\documentclass[fleqn,10pt]{wlscirep}
\usepackage[utf8]{inputenc}
\usepackage[T1]{fontenc}

\title{The network epidemiology of an Ebola epidemic}

\author[1,2,3,4,*]{Laurent H\'{e}bert-Dufresne}
\author[2,3,5]{Jean-Gabriel Young}
\author[6]{Jamie Bedson}
\author[7,4]{Laura A. Skrip}
\author[8]{Danielle Pedi}
\author[9]{Mohamed F. Jalloh}
\author[3]{Bastian Raulier}
\author[3]{Olivier Lapointe-Gagné}
\author[10]{Amara Jambai}
\author[2,3,11]{Antoine Allard}
\author[4,12,13,$\dagger$]{Benjamin M. Althouse}

\affil[1]{Department of Computer Science, University of Vermont, Burlington VT, USA}
\affil[2]{Vermont Complex Systems Center, University of Vermont, Burlington VT, USA}
\affil[3]{D\'epartement de physique, de g\'enie physique et d'optique, Universit\'e Laval, Qu\'ebec (Qu\'ebec), Canada}
\affil[4]{Institute for Disease Modeling, Global Health, Bill \& Melinda Gates Foundation, Seattle WA, USA}
\affil[5]{Department of Mathematics and Statistics, University of Vermont, Burlington VT, USA}
\affil[6]{Independent Consultant, Seattle WA, USA}
\affil[7]{School of Public Health, University of Liberia, Monrovia, Liberia}
\affil[8]{Bill \& Melinda Gates Foundation, Seattle, WA USA}
\affil[9]{Division of Global Health Protection, Center for Global Health, CDC, Atlanta GA, USA}
\affil[10]{Ministry of Health and Sanitation, Freetown, Sierra Leone}
\affil[11]{Centre interdisciplinaire en mod\'elisation math\'ematique, Universit\'e Laval, Qu\'ebec (Qu\'ebec), Canada}
\affil[12]{University of Washington, Seattle WA, USA}
\affil[13]{New Mexico State University, Las Cruces NM, USA}

\affil[*]{laurent.hebert-dufresne@uvm.edu}

\affil[$\dagger$]{bma85@uw.edu}

\begin{abstract}
  Connecting the different scales of epidemic dynamics, from individuals to communities to nations, remains one of the main challenges of disease modeling.
  Here, we revisit one of the largest public health efforts deployed against a localized epidemic: the 2014-2016 Ebola Virus Disease (EVD) epidemic in Sierra Leone. 
  We leverage the data collected by the surveillance and contact tracing protocols of the Sierra Leone Ministry of Health and Sanitation, the US Centers for Disease Control and Prevention, and other responding partners to validate a network epidemiology framework connecting the population (incidence), community (local forecasts), and individual (secondary infections) scales of disease transmission.
  In doing so, we gain a better understanding of what brought the EVD epidemic to an end: Reduction of introduction in new clusters (primary cases), and not reduction in local transmission patterns (secondary infections).
  We also find that the first 90 days of the epidemic contained enough information to produce probabilistic forecasts of EVD cases; forecasts which we show are confirmed independently by both disease surveillance and contact tracing.
  Altogether, using data available two months before the start of the international support to the local response, network epidemiology could have inferred heterogeneity in local transmissions, the risk for superspreading events, and probabilistic forecasts of eventual cases per community. 
  We expect that our framework will help connect large data collection efforts with individual behavior, and help reduce uncertainty during health emergencies and emerging epidemics.
\end{abstract}

\begin{document}

\flushbottom
\maketitle
\thispagestyle{empty}

Infectious diseases are constant yet often unpredictable threats. The 2014-2016 EVD epidemic in West Africa alone took over ten thousand lives and led to aid distributions nearing \$10 billion from 70 countries~\cite{abramowitz2017epidemics}. 
There were two major contributing factors: (i) local health systems were ill-prepared, with Ebola-affected countries falling at least 90\% short of WHO recommendations for numbers of doctors and nurses per capita~\cite{robinson2019primary}; and (ii) the declaration of the Public Health Emergency of International Concern came late: over 4 months after the first international transmission event~\cite{UNnews.2014}. 
This last systemic failure likely reflects the fact that EVD outbreaks, like most emerging outbreaks, are highly unpredictable and variable. 
Before 2014, recent Ebola outbreaks occurred in the Democratic Republic of Congo in 2014 and 2012, South Sudan in 2004, Gabon in 2001, and notably in Uganda in 2012 and 2007, but these never exceeded 500 geographically contained cases~\cite{mylne2014comprehensive}. In stark contrast, there were nearly 30,000 cases over 10 countries during the West African epidemic with widespread transmission in Guinea, Liberia and Sierra Leone, and other cases in Italy, Mali, Nigeria, Senegal, Spain, the United Kingdom, and the United States.

The spread and impact of EVD spread is determined not only by the biology of the virus but also in large part by societal and behavioral factors~\cite{skrip2020unmet, bedson2021review}. 
Consequently, in the country that reported the most number of cases and deaths, Sierra Leone, the intervention aimed to alleviate disease burden on a strained healthcare system focused on supportive care~\cite{lamontagne2018evidence} by investing significantly in behavioral interventions ranging from community engagement~\cite{bedson2020community} to traditional contact tracing~\cite{world2014contact}. 
As we will see, the latter revealed the variability of Ebola transmissions, suggesting that less than 5\% of Ebola cases caused more than one secondary infection over the course of the epidemic, yet some individuals infected dozens of others. 
We can now leverage the data collected in these efforts to develop the next generation of epidemic forecasting tools through a first study of the unique ethical and anonymized Sierra Leone Ebola Database (SLED), which consolidates data from the Government of Sierra Leone and a range of responding partners~\cite{gorina2020ensuring, agnihotri2021building}.

Here, we present case and contact tracing data from SLED. We first show that distributions of secondary infections (or transmissions) per case did not change significantly over time and that reduction in the spread of EVD was likely due to a reduction in its introduction in new communities. To validate these data, we then show that network epidemiology models informed by contact-tracing data can provide probabilistic forecasts of epidemics at the village level, capturing the heterogeneity of disease spread as well as its randomness.
This network approach contrasts with the traditional deterministic forecasts from conventional compartmental models whose only uncertainty stems from parameter confidence rather than from the intrinsic randomness of epidemics.
Importantly, we also show how to infer valuable individual-level contact statistics from population-level incidence data. 
Our network epidemiology framework helps us better understand the dynamics of EVD in Sierra Leone. 
It also opens the way to multi-scale inference of epidemic dynamics, connecting population incidence, probabilistic forecasts in communities, and individual-level heterogeneity. 

One key obstacle that precluded previous validation of models accounting for individual heterogeneity is the rarity of multi-scale data.
Most epidemic data are coarse-grained over large populations (e.g. nations, states, etc.), while models often require individual level data (e.g. distribution of secondary infections or contact patterns) to provide forecasts at the level of communities (e.g., villages or towns).
In that respect, the large contact tracing and data collection efforts accomplished during the EVD epidemic in Sierra Leone are unique in their scale and quality.

\begin{figure}
    \centering
    \includegraphics[width=0.54\linewidth]{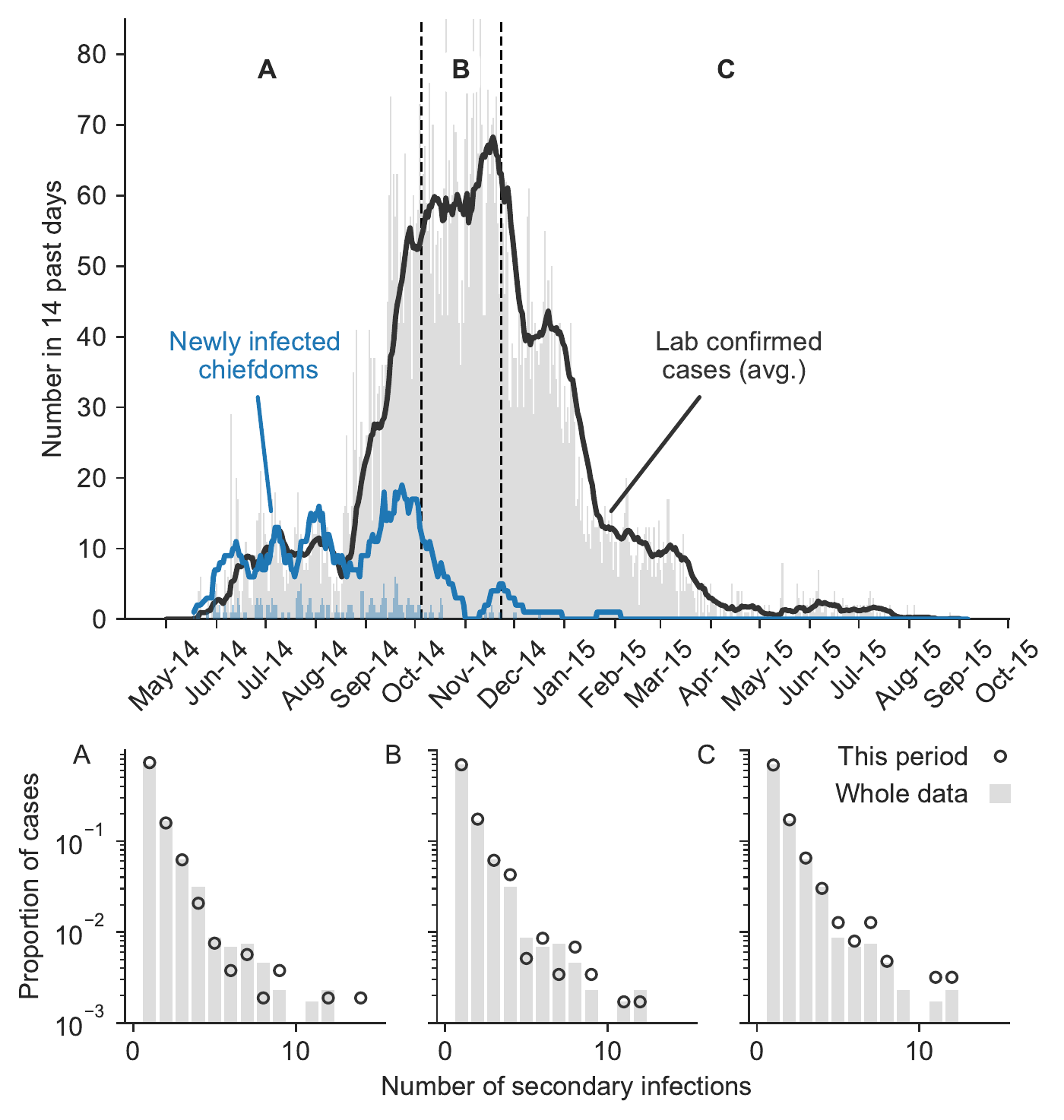}
    \caption{\textbf{Incidence data of EVD in Sierra Leone.} The top panel shows confirmed cases of EVD in Sierra Leone (grey bars, with running 7-day average in grey line) and new number of chiefdoms reached (14-day average in blue line, data as reported in Fang et al.~\cite{fang2016transmission}) between May 2014 and September 2015. Chiefdoms are administrative units in Sierra Leone, typically consisting of roughly $100$ villages. By the end of the epidemic, EVD had reached 113 (or about 75\%) of the 149 chiefdoms present at the time. Labels \textbf{A},~\textbf{B}~\&~\textbf{C} correspond to secondary case distributions shown in the bottom panels and compiled from the contact tracing data in SLED. Dates were chosen to have roughly equal numbers of reported cases in each time period and to correspond to the rise, peak and decay of the epidemic.
    Our framework allows us to connect these different scales of data: Macroscopic incidence curve across the entire country, mesoscopic forecasts per village, and secondary infections per individual case.
    This framework supports the data shown in panels \textbf{A},~\textbf{B}~\&~\textbf{C} and suggests that the reduction in introduction in new regions by the end of September 2014 was the main driver in reducing incidence of EVD almost two months later. 
    }
    \label{fig:timeline}
\end{figure}

During the West African EVD epidemic, contact tracing efforts were sustained to varying degree but spanned the entire epidemic~\cite{team2014ebola}, see Fig.~\ref{fig:timeline}. 
These efforts followed the guidelines of the World Health Organization on how to establish and conduct contact tracing during filovirus disease outbreaks~\cite{world2014contact}. 
Therein, contact tracing is defined as the identification and follow-up of persons who may have come into contact with an infected individual. 
Identification is done by investigating high-risk categories of contacts, which include but are not limited to individuals who: (i) touched an infected patient's body fluids (blood, vomit, saliva, urine, faeces), (ii) had direct physical contact with the body of a patient (alive or dead), (iii) touched or cleaned the linens or clothes of a patient, (iv) were laboratory workers who had direct contact with specimens collected from suspected Ebola patients without appropriate infection prevention and control measures, or (v) were patients who received care in a hospital where EVD patients were treated before the initiation of strict isolation and infection prevention and control measures.

For disease modeling and forecasting, several key insights can be extracted from contact tracing data ~\cite{althouse2020superspreading}: Identifying the critical types of contacts, settings and mechanisms that promote transmission, inferring the frequency of superspreading events, and so on.
The distribution of secondary infections identified through contact tracing around a given index case also reflects the heterogeneity of disease spread~\cite{blumberg2013inference, hebert-dufresne2020r0b}.
In particular, the average number of secondary infections directly reflects the basic reproduction number $R_0$ of the epidemic, which is often the sole focus of disease models ignoring the richness and heterogeneity of the underlying distribution.

For the EVD epidemic in Sierra Leone, the distribution of secondary infections collected through contact tracing is shown for three key periods of the epidemic in Fig.~\ref{fig:timeline}.
First, we note the absence of data regarding cases that did not cause any secondary infections.
This is a direct shortcoming of the data collection which did not distinguish between cases where contact tracing efforts did not identify secondary infections and cases where there were no contact tracing efforts whatsoever.

Second, the observed distributions show significant heterogeneity, stressing the need to account for more than simply their average (e.g., $R_0$).
In individuals at the source of at least one transmission event, we find an average of roughly 1.6 secondary infections in all three periods (1.624, 1.622, and 1.660 respectively).    
Yet, out of cases that do transmit the disease, 85\% lead to a single secondary infection while at least 1 in a thousand cases leads to 10 infections or more.
It is therefore dangerous to assume that the EVD epidemic was driven by average individuals.
The majority of cases create fewer secondary infections than the average, while a minority create an order of magnitude more infections than average~\cite{lloyd-smith2005superspreading,hebert-dufresne2020r0b}.

Third, and perhaps most strikingly, the distribution of secondary infections do not statistically-significantly change across the three observed periods.
These periods --- May through September 2014, October to mid-November 2014, and onward --- were chosen to contain similar numbers of contact tracing data but also roughly correspond to the onset of the epidemic peak, the epidemic peak itself, and the decay of the epidemic.
One could reasonably assume that the large public health efforts which helped limit the spread of EVD would here be reflected as a decrease in the frequency of higher numbers of secondary infections.
However, while public health efforts did reduce mortality 75 to 40\%~\cite{lamontagne2018evidence,lamontagne2019evolution}, they did not appear to influence the distribution of secondary infections as observed through contact tracing.
This is a surprising result considering the scale of epidemic control efforts. 

We thus aim to validate this distribution of secondary infections.
Are these contact tracing data reliable enough to inform epidemic models and forecasts? The contact tracing data is of course not perfect, with probable biases around the fact that contacts can be missed and pressure to cover super-spreading events. The latter bias may have increased over time, cancelling out changes in the distributions of secondary infections. Conversely, the data may be accurate and the epidemic might have been brought to an end by an earlier reduction of the rate of introduction of EVD in new communities (see blue curve in the main panel of Fig.~\ref{fig:timeline}) rather than a direct reduction in secondary infections. Our understanding of what stopped the epidemic therefore rely critically on being able to validate the observed contact tracing data.

We here develop a framework for network epidemiology starting directly from contact tracing data rather than contact patterns or incidence rates~\cite{hebert-dufresne2020r0b}.
We define transmission trees by their distribution of secondary infections; i.e., the distribution of the number of infections (transmissions) caused by individuals (nodes in the tree).
Once a distribution is specified, network theory allows us to forecast outbreak sizes by calculating the size distribution of the resulting infection trees.

The calculation is standard and can be summarized as follows\cite{newman2001random}.
Given the case distribution taken from the contact tracing data, we define $G_1(x)$ as the probability generating function (PGF~\cite{wilf2005generatingfunctionology, miller2018primer}) of the secondary infection distribution $\lbrace u_n \rbrace$, i.e.~\cite{newman2002spread}
\begin{equation}
  G_1(x) = \sum _{n=0}^{\infty} u_n x^n \; ,
  \label{eq:g1}
\end{equation}
where $u_n$ corresponds to the fraction of individuals (or \textit{nodes} in the transmission network) who caused $n$ secondary infections and $x$ is a counting variable for these secondary cases. 

Since the network in question is the network of actual transmissions, $G_1(x)$ generates the number of secondary infections whose average is the basic reproduction number $R_0 = \sum_{n=0}^\infty n u_n = dG_1(x)/dx\vert_{x=1}$.
Thus, contact tracing provides the distribution generated by $G_1(x)$ rather than the usual degree distribution~\cite{newman2001random, newman2002spread} of networks generated by some other PGF $G_0(x)$.

In fact, since an individual at one end of a random transmission event is $n$ times more likely to be involved in $n$ transmission events, we can recover the PGF associated with the unobserved degree distribution, $G_0(x)$, from $G_1(x)$ like so~\cite{hebert-dufresne2020r0b}:
\begin{equation}
  G_0(x) = p_0 + (1 - p_0) \left[ A \int G_1(x) dx + B \right]
         = p_0 + (1 - p_0) \left[\sum _{n=1}^{\infty} \frac{u_{n-1}}{n\ \left[\sum _{n'=1}^{\infty} u_{n'-1}/n'\right]} x^n\right] \; ,
\end{equation}
where the constants $A$ and $B$ are chosen such that $G_0(1) = 1$ and $G_0(0) = p_0$, where $p_0$ is the probability that the primary case of a new cluster does not transmit the disease at all.

Furthermore, given that links represent actual transmissions, we can select a random primary case and know that all nodes that can be reached from that node will be infected.
In other words, the size distribution of connected components is equivalent to the distribution of final outbreak size in a given population.
To calculate this distribution, we define $H_1(x)$ as the PGF for the distribution of the number of nodes reachable by following a random transmission~\cite{newman2001random, newman2002spread, newman2007component}. 
At the end of this random transmission will be one node with secondary infections generated by $G_1(x)$ and each of these transmissions will in turn also reach a number of nodes generated by $H_1(x)$. 
The PGF $H_1(x)$ must therefore satisfy the following self-consistency condition (or recursion):
\begin{equation} \label{eq:h1}
  H_1(x) = xG_1(H_1(x))
\end{equation}
where $G_1(H_1(x))$ sums the number of nodes reachable from the secondary infections of a given node, and the $x$ factor counts the node itself.
Similarly, the size distribution of outbreaks starting from a random node (or primary case) is generated by
\begin{equation} \label{eq:h0}
  H_0(x) = xG_0(H_1(x)) \; .
\end{equation}
where we replace $G_1(x)$ in Eq.~(\ref{eq:h1}) by $G_0(x)$ to start from a random node rather than a random transmission event.

This framework yields a probabilistic forecast for the sizes of outbreaks that eventually die out, i.e. the vast majority of outbreaks. Indeed, Eqs. \eqref{eq:g1}--\eqref{eq:h0} connect the distribution of secondary infections, $G_1(x)$, with the distribution of cases in a community, i.e., the outbreak size distribution, $H_0(x)$.
We have designed a Bayesian framework, detailed in our Supplementary Information document, that implements this relation and allows us to move freely from one point of view to the other while accounting for our prior knowledge on epidemic parameters (e.g. $R_0$ and dispersion $k$). 
In other words, our Bayesian framework can either predict a distribution of outbreak sizes based on a distribution of secondary infections gathered from contact tracing (\textit{forward prediction}), or inversely predict individual heterogeneity based on a distribution of cases per community (\textit{backward prediction}).

\begin{figure}
    \centering
    \includegraphics[width=0.7\linewidth]{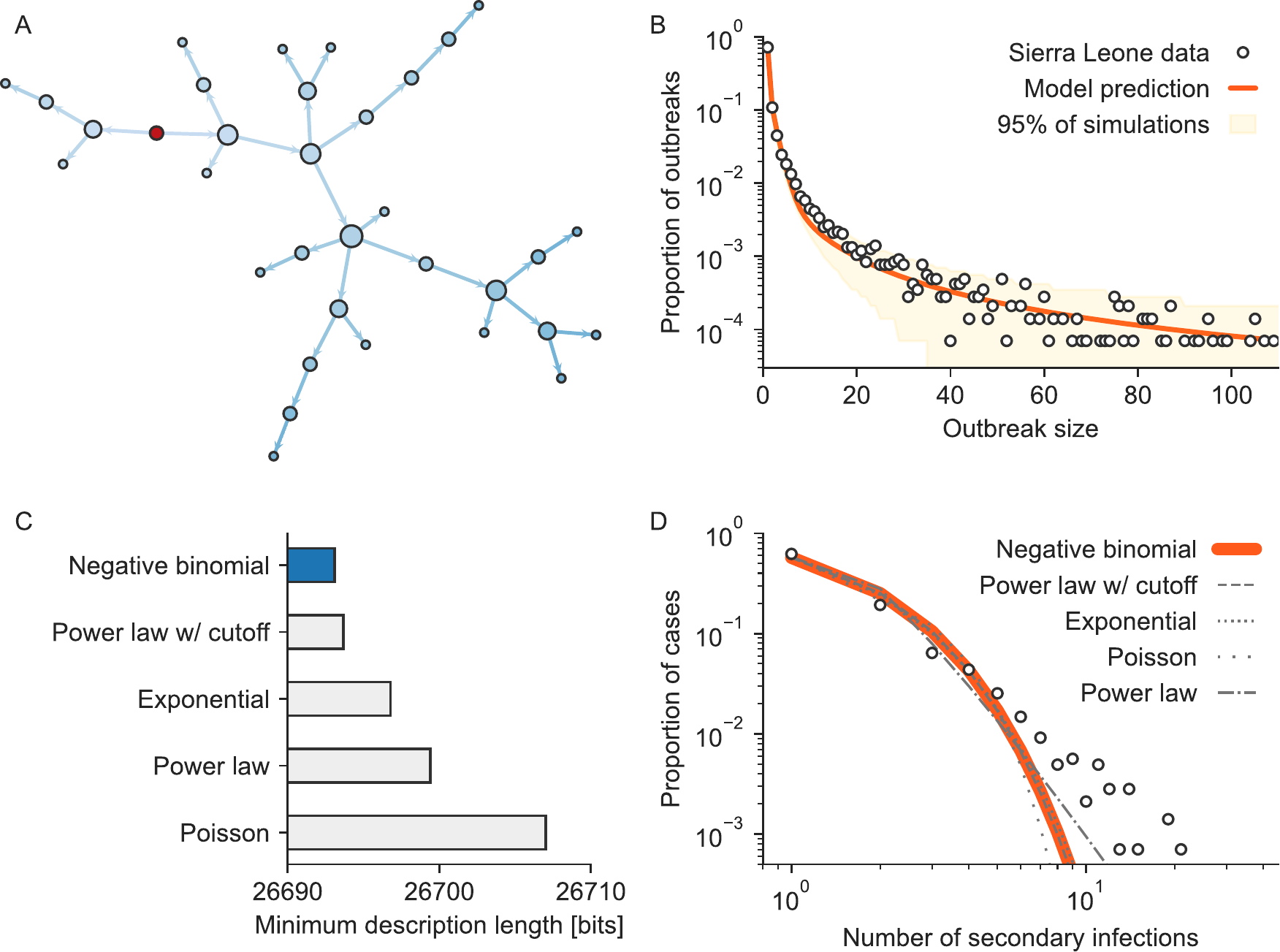}
\caption{
{\bf Mapping the distribution of cases per community to the contact distribution.}  Network epidemiology provides a framework to predict the distribution of cases in a community or cluster based on its contact distribution by predicting sizes of transmission trees (A-B). \textit{Vice versa}, we can ask what contact distribution should we expect given a distribution of cases in a community (C-D). 
\textbf{(B)} When we use the measured distribution of contacts per infectious individuals we are able to make a \emph{forward prediction} of the final distribution of transmission trees accurately, even when we add errors to the measurements. This matches the distribution of cases per village over the entire EVD epidemic in Sierra Leone as reported in the SLED. \textbf{(C)} To make \emph{backward predictions}, we test several models of contact distributions and optimize their parameters by fitting a Bayesian model to the distribution of outbreaks. 
We compare models through their minimum description length (MDL), where a lower MDL implies a better fit by a more parsimonious model.  \textbf{(D)} More importantly, this approach can help recover statistics of the underlying contact distribution even when our framework only had access to the outbreak size distribution. All models have similar MDLs and are plausible alternatives, some of them aligning quite closely with each other. The negative binomial model offers the best fit by a small margin. All models provide good fits of the bulk of the observed distribution of secondary infections as reported in the SLED.}
    \label{fig:results}
\end{figure}

Figure~\ref{fig:results} shows the result of using our approach for forward and backward predictions using a Bayesian model applied to contact tracing and village-level case data from the SLED.
The forward prediction only involves fitting the frequency of cases leading to no secondary infections, as per the data collection shortcoming explained above.
Figure~\ref{fig:results}(B) shows that forward prediction of outbreak sizes provides accurate forecasts of the final distribution of cases per village as reported in SLED (a very small Jensen-Shannon divergence of $0.005$).
For backward prediction, a large family of potential transmission distributions is investigated: the Poisson distribution assumed by usual well-mixed models~\cite{Kermack1927, diekmann1995legacy, hebert-dufresne2020r0b}, a discrete exponential distribution~\cite{newman2001random}, as well as more realistic negative binomial~\cite{lloyd-smith2005superspreading} and power-law distributions~\cite{clauset2009powerlaw}. 
For all the models, we infer the parameters of the distribution with a Bayesian method and also consider flexible versions of the models where $u_0,...,u_{d-1}$ are free parameters.
We then apply a model selection framework based on the minimum description length~\cite{grunwald2007minimum, mackay2003information} to select which contact distribution provides the most parsimonious model. 
Figure~\ref{fig:results}(D) shows that this methodology is consistent, meaning that we can recover the bulk of a contact distribution from its outbreak size distribution. 
Furthermore, as shown in Fig.~\ref{fig:results}, the contact distribution predicted by our Bayesian model is close to the observed one.

This preliminary analysis of the data from the EVD epidemic in Sierra Leone is promising on two fronts. 
It constitutes a validation both of the theoretical foundations of network epidemiology and of the value of contact tracing data. 
One key assumption made by this current model is that the distribution of secondary infections does not change through time, which was motivated by our original observations in Fig.~\ref{fig:timeline}. If this assumption did not hold, contact tracing data would be required to define different PGFs $G_1(x,t)$ for different periods of time $t$ to capture time evolution~\cite{noel2009time}. Another less obvious assumption comes from our comparison with empirical data, where we assume that villages in Sierra Leone, as defined in the SLED, are the characteristic scale of the epidemic. The strong fit of the model prediction to the distribution of outbreak sizes implies that most transmission trees are confined within a given community, and that EVD outbreaks within villages are effectively independent realizations of the same stochastic process. 

Importantly, the analysis presented in Fig.~\ref{fig:results} stems in large part from the laborious and thorough contact tracing efforts in Sierra Leone -- perhaps more so than the model itself. 
Early modeling of new outbreaks often focuses solely on the average growth of publicly available time series ascertain the basic reproduction number $R_0$, which can inform deterministic models for the exponential growth of the outbreak. Our modeling framework suggests that knowing the variability of new cases reported frequently could inform us about the heterogeneity of secondary infections. This information could then be used to provide probabilistic forecasts to supplement the traditional epidemic growth model.

We thus broaden this framework and aim to infer the distribution of secondary infections from population incidence data to provide probabilistic community forecasts from early epidemic data. 
To do so, we rely on a negative binomial distribution as a candidate for $G_1(x)$, following both our model selection framework and existing literature~\cite{lloyd-smith2005superspreading}.
To account for temporal delays between cases we rely on Approximate Bayesian Computation (ABC)~\cite{sunnaker2013approximate, riou2020pattern} rather than an explicit likelihood, and simulate the network of infections as it grows.
We randomly generate artificial time series of cases over a 90-day period starting from the first confirmed EVD case.
For every new case, we sample the candidate distribution of secondary infections and, for every secondary case, we then sample a gamma distribution of generation time to account for temporal delays between infections and case detection.
The joint posterior distribution for the parameters of these candidates are then approximately sampled based on a simple error metric: At six early sampling points -- 12, 30, 45, 60, 75 and 90 days -- we calculate and sum the difference in cases between the generated time series and the actual number of confirmed and suspected cases (we vary the number of sampling points in the Supplementary Information).
In the limit of a vanishingly small error threshold to accept a sample of parameters, we know that the generated approximate posterior distributions will be equivalent to the true Bayesian posteriors.
Tuning this error threshold, can balance the speed of the approach with the desired level of approximation to the true posteriors.

\begin{figure}
    \centering
    \includegraphics[width=\linewidth]{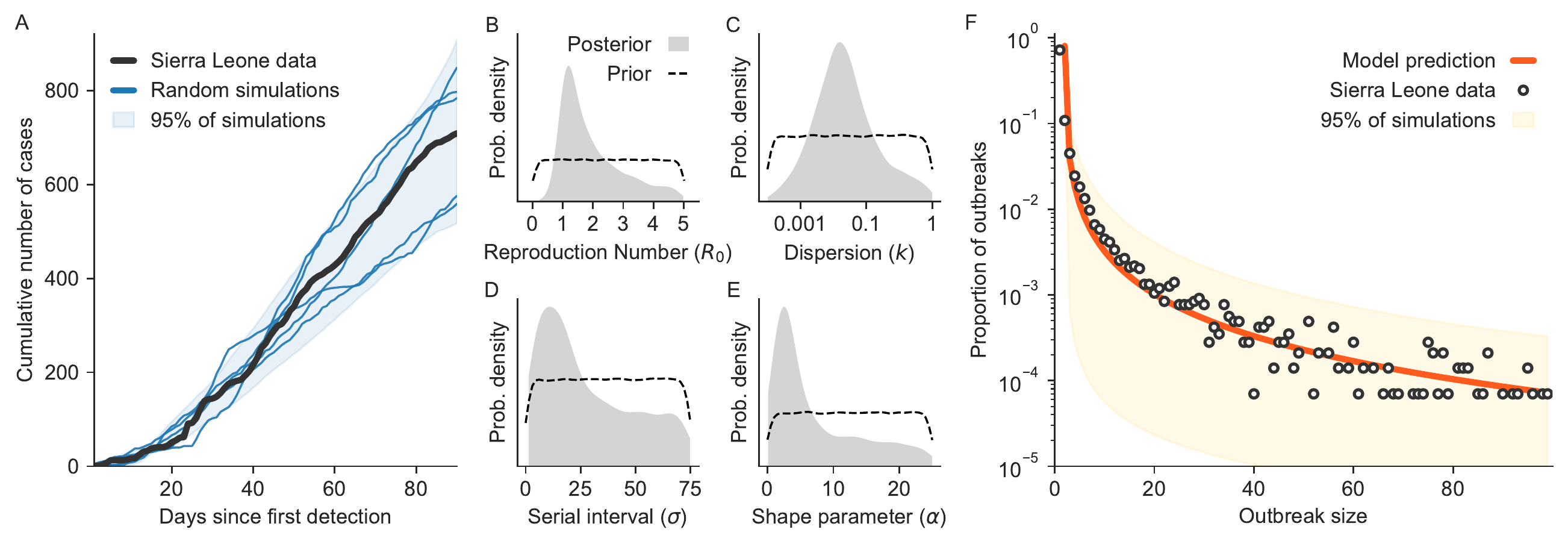}
    \caption{\textbf{Inferring individual and community level transmission from population case data.}
    (A) The time series of cumulative EVD cases in Sierra Leone (black thick line) along a sample of the randomly generated time series used to computed the posterior distribution.  Each of the accepted sequence deviates less than 30\% from the Sierra Leone time series on days 15, 30, 45, 60, 75 and 90.
    (B--E) Comparison between the non-informative prior distributions and the marginal posterior distributions for parameters $R_0$, $k$, $\sigma$ and $\alpha$ obtained from the ABC procedure.
    The mode of the posterior distributions are: $R_0: 1.01$, $k: 0.00042$, $\sigma: 2.8$, and $\alpha: 2.57$.
    (F) Comparison between the posterior outbreak size distribution predicted by the PGF formalism using the posterior distribution inferred with the ABC procedure and the outbreak size distribution observed in Sierra Leone during the 2014-2016 EVD epidemic.
    A detailed description of the ABC procedure and of the PGF formalism is provided in the Supplementary Information document.}
    \label{fig:ABC}
\end{figure}

Figure~\ref{fig:ABC} presents the results obtained from this procedure.
We start with uniform prior distributions over ranges for the four key parameters of our framework: (i) The average number of secondary infections ($R_0$), the dispersion coefficient ($k$) of the underlying negative binomial distribution of secondary infections, the average serial interval ($\sigma$) between epidemic generations, and a shape parameter ($\alpha$) for the underlying gamma distribution.
Figures~\ref{fig:ABC}B--E compare these non-informative priors with the well peaked marginal posterior distributions and confirm that the early days of the EVD epidemic contained enough useful information to inform the ABC sampler.
More importantly, the inferred parameters are consistent with the contact tracing data that would come later in the epidemic (i.e., $R_0$ around 1.6) and sufficient to provide a posterior distributions of probabilistic forecasts consistent with the final distribution of cases per village, see Fig.~\ref{fig:ABC}F.
In short, this network epidemiology framework connects the population data available over the first 90 days of the epidemic with the large final community-level data of cases per village, and does so through a parametrization of key epidemiological parameters including secondary infections at the individual level.

Altogether, our results show how the individual-, community- and population-scale of epidemic dynamics can be integrated under network epidemiology. For the EVD epidemic in Sierra Leone, these three scales correspond to the number of early cases over time in the country (population), the distribution of outbreak sizes per villages (community) and statistics regarding the distribution of secondary infections (individual). Using a single scale of data, this framework allows us to infer or predict others and make probabilistic disease forecasts.

Given how well our data-driven model explains observed data, this validation of a network epidemiology framework also illustrates the quality of the data collected during the unprecedented contact tracing and disease surveillance efforts, despite its challenges, of the 2014-2016 EVD epidemic in Sierra Leone~\cite{mcnamara2016ebola}. The framework as a whole could not work without both strong theoretical grounding and high quality data. The results therefore highlight the importance of efforts to strengthen the Epi Info VHF application used in West Africa and the potential and need for data standardization and sharing in support of comprehensive collection and collation in real time~\cite{schafer2016epi} such as the Go.Data application~\cite{world2020go}. 
Importantly, we note that several of these data were potentially available in real time~\cite{international2016exposure} but not necessarily to the wider public. 
Wider access to anonymous summary statistics from quality contact tracing could be leveraged in modeling and forecasting to better guide interventions at the beginning and throughout an epidemic.

Our mathematical framework consists mostly of simple operations over polynomial functions -- composition or convolution of PGFs -- which require very limited computational capacity. 
The code based provided with this paper can therefore be used on any personal computer to provide real-time probabilistic forecasts or to infer epidemiological parameters as new data emerge.
In particular, the probabilistic forecasts provide a direct risk assessment of outbreak sizes which could be conditioned on the number of cases to date in a community.
These assessments could therefore help guide the prioritization of preventative interventions such as risk communication and community engagement~\cite{bedson2020community} or vaccination~\cite{wang2016statistical} in particular communities targeted based on local forecasts from contact tracing data or case incidence.

Future work will aim to expand this framework to emerging epidemics through both revamped recommendations for contact tracing protocols~\cite{world2020contact} and a more general mathematical model. On the logistics front, distinguishing null from missing data to identify the frequency of cases leading to zero secondary infections would help reduce uncertainty in our forecasts. A more general approach to anonymized contact tracing data could therefore look like an anonymized multidimensional distribution over classes of cases (e.g. children and adults) tracking their number of contacts ($c$) identified, number of contacts reached ($r\leq c$), and number of positives cases identified ($n\leq r$). On the mathematical front, this would allow for a more general model where we could consider contact networks rather than explicit transmission trees to account for heterogeneous contact patterns and transmissibility over that network~\cite{meyers2007contact, allard2009heterogeneous, allard2017asymmetric}.

Implementation of this stochastic mathematical model can go hand in hand with data collection in future emerging epidemics. It offers real-time estimates of key epidemic parameters such as $R_0$ and the serial interval, expectations for the dispersion of secondary infections, probabilistic forecasts when an outbreak occurs in a given community, and much more. Moreover, the Bayesian posteriors help us quantify how much uncertainty exists around specific parameters, guiding additional data collection. Together with contact tracing and disease surveillance, this framework will help us separate the impact of data uncertainty from the inherent randomness of emerging epidemics.

\section*{Acknowledgements}
This work was supported by the 'Data Modeling Community Engagement in Health Emergencies' project funded by the Bill \& Melinda Gates Foundation, including contributions from the Institute for Disease Modeling. The Sierra Leone Ebola Database (SLED) data utilized in this paper were accessed as a component of this project. The authors wish to thank the Ministry of Health and Sanitation of the Government of Sierra Leone, the SLED team in Sierra Leone and the US Centers for Disease Control (in particular Dr. Yelena Gorina, Negasi Beyene and John Redd) for their support in reviewing the project and facilitating data access. LHD, JB, LS, DP and BMA were supported by Bill \& Melinda Gates through the Global Good Fund. LHD and JGY acknowledge support from the National Institutes of Health 1P20 GM125498-01 Centers of Biomedical Research Excellence Award. AA acknowledges financial support from the Sentinelle Nord initiative of the Canada First Research Excellence Fund and from the Natural Sciences and Engineering Research Council of Canada (project 2019-05183). The funders had no role in study design, data collection and analysis, decision to publish, or preparation of the manuscript.

\section*{Additional information}
\textbf{Competing financial interests} The authors declare that they have no
competing financial interests.

\end{document}